Preprint submitted to Acta Astronautica

# Beyond the Beacon: State and Corporate Competition after an Information-Rich Technosignature Detection

Michael A. Garrett [a,b,c,*]

[a] *Jodrell Bank Centre for Astrophysics, Department of Physics and Astronomy, The University of Manchester, Oxford Road, Manchester M13 9PL, United Kingdom*
[b] *Leiden Observatory, Leiden University, P.O. Box 9513, NL-2300 RA Leiden, The Netherlands*
[c] *University of Malta, Institute of Space Sciences and Astronomy, Msida, MSD2080, Malta*

* Corresponding author. E-mail address: michael.garrett@manchester.ac.uk

A B S T R A C T

The 2026 update to the Declaration of Principles Concerning the Conduct of the Search for Extraterrestrial Intelligence (SETI) provides a forward-looking framework for post-detection practice. Like earlier versions, it addresses information content directly: a confirmation report should record any detected content, and the underlying data, methods, and code should be openly preserved and disseminated. The declaration is necessarily concise and is intended as a living instrument, to be supported by supplementary material that examines specific issues in greater depth. This paper contributes to that process by considering the possible detection of an information-rich technosignature: a signal carrying recoverable structure, data, imagery, or machine-readable content. Whereas the prevailing concept of a detection remains that of a beacon, whose content is exhausted by demonstrating that a transmitter exists, a signal with recoverable and potentially exploitable content would shift much of the post-detection challenge from verification to information stewardship, while encouraging competitive efforts to decode and use the signal. Few studies have addressed this scenario directly, and the geopolitical literature has so far framed post-detection competition mainly in terms of state actors. I argue that this framing is no longer sufficient. Large corporations now possess the computing power, AI capabilities, and, in some cases, receiving infrastructure needed to independently pursue an extraterrestrial signal without any obligation to disclose their findings. This largely unexamined dimension materially changes the post-detection landscape. These considerations bear directly on the declaration's commitment to openness and transparency, and on how that commitment can be upheld when powerful actors outside the SETI community may not share it. I identify information-rich technosignatures as a topic for supplementary guidance and offer a preliminary set of questions for the community to consider.

## 1. Introduction

The detection of a signal of unambiguously extraterrestrial and artificial origin would rank among the most consequential events in the history of science. In anticipation of such a detection, the SETI community has, over many decades, generated a body of best practice, embodied in the "post-detection protocols", intended to guide researchers in the period following a candidate detection. These protocols, articulated in the Declaration of Principles Concerning the Conduct of the Search for Extraterrestrial Intelligence and periodically revised, most recently in the 2026 update [1], outline a well-defined sequence: the preservation of the observational record, the exclusion of natural and anthropogenic explanations, the securing of independent verification, and the open and responsible communication of a confirmed result. The 2026 revision broadens the scope from the historical radio-signal scenario to technosignatures more generally. It strengthens guidance on data handling, spectrum protection, and communication in a contemporary media environment. It also proposes the establishment of a standing post-detection committee with expertise spanning science, ethics, law, the social sciences, and communication [1,2].

An important feature of the declaration, for present purposes, is that it addresses information content directly, as did the 2010 revision before it. It requires that a confirmation report record any detected information content and encourages the preservation and dissemination of the data, analysis methods, and code underlying a detection. Archiving in more than one geographically separated repository is recommended so that results can be independently reproduced and secured [1]. The possibility that a signal might contain recoverable content is thus acknowledged within the framework, and the response it commends - to document the content and to make the underlying material openly available for scrutiny - is consistent with the broader and well-founded norms of open science. Equally important, the declaration is not conceived as a closed or final text. It is a living instrument, designed to be complemented by supplementary best practices as the topic develops [1,2].

A declaration of principles is necessarily concise. It establishes the norms and the sequence that should govern any detection, but it cannot, and does not set out to, treat in operational detail the many distinct circumstances that a real detection might present. This is by design: the framework is intended to be extended through supplementary contributions, each addressing a particular aspect of post-detection practice in a depth that a short declaration cannot. Several such aspects will warrant dedicated attention. Among them is the personal safety and well-being of the individual researchers who make and announce a discovery, and who would no doubt face intense scrutiny from the public, the media, and the scientific community. Others include the finer details of data custody and communication, and, of course, the question of whether any response to a confirmed detection should be made. This work is already under way. Draft best-practice guidelines have recently been presented on science and risk communication for technosignature searches and detections [3], and on the safety of scientists engaged in technosignature and biosignature research [4]. Each takes up an issue that the 2026 declaration includes but does not develop in great detail. It is to be expected, and indeed hoped, that supplementary contributions will in due course address the remaining issues. The present paper takes up one further aspect, and one that I personally regard as among the most pressing: the handling of information content, and the distinctive demands that would arise if a detected signal proved to carry recoverable and potentially exploitable material.

It is in that spirit of continuing development of the protocols that this paper is offered. Its purpose is to draw attention to one scenario that would repay dedicated consideration as supplementary guidance is developed: the detection of what may be termed an information-rich technosignature, defined here as a candidate signal for which the received data support the presence of recoverable structure, data, imagery, or machine-readable content. The motivation is straightforward. While the declaration provides for information content, the conception of a detection that most readily informs discussion, in the wider community and in much of the surrounding culture, remains that of a beacon: a cosmic lighthouse whose message is nothing more than a demonstration that a transmitting civilisation exists. For a beacon, the requirement to record information content is satisfied trivially, and the encouragement to preserve and openly disseminate the data carries no obvious difficulty, because the data contain nothing that could be exploited or misused. A detection whose content is recoverable and potentially valuable is a materially different case, and one that supplementary guidance would be well placed to consider in advance.

The remainder of this paper develops that observation. It considers the information-rich scenario in relation to the existing declaration and sets out why such a detection would shift much of the post-detection effort from verification towards the stewardship of information. It describes the competitive pressures that such a detection is likely to generate, on the part of both state and private commercial actors, and asks how the declaration's commitment to openness might best be upheld under those pressures. The paper seeks only to identify a topic worthy of attention, and to offer a preliminary set of questions, as the community develops supplementary contributions in support of the framework it has established.

## 2. Information content in the current framework and literature

The treatment of information content in the current framework is best appreciated against the background of the scholarship on which the field draws, and which supplementary guidance would build upon. Several areas of post-detection research bear on the information-rich scenario, and it is useful to identify what each contributes.

The most directly relevant area is the tradition concerned with deciphering a received message. Earlier work on a post-detection decipherment matrix proceeds explicitly from the premise that a detected signal may possess decipherable linguistic or structural content and prepares the analytical means by which such content might be recovered [5]. The most developed precursor to the present discussion is the "DISC Quotient" (Decipherment Impact of a Signal's Content): a scheme that introduces a standardised scale to quantify humanity's current progress in understanding a potential alien signal [6]. Its emphasis lies on characterising and sequencing decipherment - the complementary questions of who may access recoverable content, on what terms, and how openness is sustained when that content has value to competing actors are not addressed. A recent and instructive demonstration of the associated challenges is the *A Sign in Space* project, in which a simulated extraterrestrial message was transmitted and released for public decoding; a distributed community of thousands extracted the signal within days, yet decoding its embedded image - a molecular diagram of amino acids - took a further year to be uncovered [7]. The exercise illustrates, in miniature, the staged character of the task: receipt, detection, information extraction, decoding, and interpretation. It also demonstrates that a real detection would provoke an impressive level of interest and response that is directly relevant to the discussion that follows. The same exercise also offers an insight into how openly released data can attract interest beyond the community for which they were intended, including commercial interest that the project took steps to pre-empt [8], a point further developed in Section 4.

Another area concerns readiness and societal impact, and provides much of the interdisciplinary infrastructure within which supplementary guidance would be prepared. Denning and colleagues have set out a post-detection research roadmap spanning evidence of engineering or message content, societal impact, communication, protocols, and interdisciplinary coordination [9]; the establishment of the SETI Post-Detection Hub and more recently the SETI Institute's Discovery and Futures Lab seek to build the interdisciplinary capacity a detection would demand [10]; and recent work has set out directions for technosignature research and readiness across many different dimensions [11]. Within this programme, scenario-based and strategic-foresight methods have been developed specifically to explore post-detection futures, combining a range of plausible detection situations with contrasting long-term contextual backdrops against which a detection might unfold [11,12,13]. These tools are well suited to examining the information-rich scenario discussed here, including the likely behaviour of state and commercial actors under different future conditions. The broader question of the societal impact of a discovery has a substantial literature of its own [14]. This work treats decoding, societal impact, and infrastructure as elements of a general readiness agenda. The present paper isolates one element within that agenda - the distinction between a signal whose content is null and one whose content is recoverable and potentially exploitable. In particular, it identifies the handling of data as a priority for dedicated supplementary guidance.

The geopolitics of a detection has also received direct attention and bears immediately on the discussion below. Wisian and Traphagan have argued that even a passive detection carries an underexplored risk that state-level actors might seek an information monopoly over contact and communication with an extraterrestrial intelligence, and have recommended that SETI facilities adopt security measures to forestall this possibility [15]. Wright, Haramia, and Swiney have responded that monopolising communication with an extraterrestrial intelligence is technically and politically unlikely, and that transparency and data sharing, rather than security and secrecy, offer better protection, alongside further development of post-detection protocols and education for policymakers [16]. That SETI has never been insulated from geopolitics is, moreover, a matter of historical record [17,18]. This exchange establishes that the competitive dimensions of a detection are already recognised in the literature, and that the community continues to weigh how the principle of openness is best applied in their presence, a deliberation to which the information-rich scenario is directly pertinent. Legal scholarship has also addressed the consequences of contact, including the applicability of existing human law to an encounter with an extraterrestrial intelligence [19]. This bears on the question of which institutions might have authority over a detection and its aftermath, though it does not principally concern the handling of recoverable signal content.

Taken together, these areas of research provide strong foundations. The existence of information content, the staging of its decipherment, the geopolitics of a detection, and the legal context of contact are each addressed. The contribution of the present paper is to bring these threads together around a single scenario: the detection of recoverable content whose potential value gives rise to competition, and the implications this carries for post-detection policy, practice and supplementary guidance.

## 3. The distinctiveness of an information-rich detection

The difference between a beacon and an information-rich signal is not one of degree. An information-rich detection would alter the character of the post-detection situation in several respects, each of which future guidance should usefully take into account.

A simple beacon establishes a single proposition: that humanity is not alone. This is a conclusion of profound significance, but it is a fact rather than a practical resource. An information-rich signal may, by contrast, contain knowledge - representations of physical, mathematical, biological, or engineering concepts possessed by another civilisation. The mere prospect that such content might be recoverable changes the status of the detection. It becomes not only an object of scientific inquiry but also one that various actors would have strong incentives to obtain and to control. Applicable knowledge confers advantage, and the pursuit of advantage is a durable feature of the conduct of states and commercial organisations alike. Contemporary competition over advanced artificial intelligence (AI) offers a familiar illustration of how potentially transformative information technologies can attract both national and corporate strategic interest.

Even a beacon carrying no recoverable content would set off an immense and sustained scientific effort. Verification and announcement would be only the beginning: observatories across the world and across the electromagnetic spectrum would turn to the source around the clock, working to characterise the host star, planet, or platform, to search for associated emission, and to tease from the data whatever hints of the transmitting civilisation's location, capabilities, and technical attributes it might yield. This campaign could run for years, ultimately bounded not by any shortage of effort but by the limits of our own instrumentation. What distinguishes the information-rich case is not that it replaces this astronomical work - that would proceed in either scenario - but that it adds to it an undertaking of an entirely different kind. The recovery, decoding, and interpretation of structured content is also likely to be a protracted undertaking, as the decipherment literature and the "A Sign in Space" exercise both indicate [5,6,7]. Throughout this effort, the incentive to claim the lead in decoding, and to potentially act on provisional interpretations, would be considerable. Decipherment might of course prove easier than has often been supposed, particularly in the era of large language models and related AI systems. Such a possibility strengthens rather than weakens the case for guidance, since rapid partial decoding would only increase the pressure to interpret and act on uncertain results.

A practical complication is that the distinction between a beacon and an information-rich signal may not be evident at the moment of detection. The information content of a signal could be recognised progressively: first through structure that distinguishes it from a pure carrier, such as modulation, periodicity, or framing; then through the recovery of a symbol stream; and only later through decoding and interpretation. Thus a signal originally characterised as a beacon might later prove to be information-rich as sensitivity, bandwidth, or integration time improve, or as analysis matures - a narrowband carrier could, for example, accompany a weaker, broader-band component revealed only by deeper follow-up observations. Conversely, apparent structure may ultimately prove to be instrumental or terrestrial in origin. There will therefore be an interim period, potentially a long one, during which the extraterrestrial origin of a signal is being established while its information content remains uncertain. Importantly, the declaration's openness approach already applies throughout this period: independent verification requires that other facilities receive the details needed to re-observe the source, and once a confirmed detection is announced, any adequately equipped facility can record the signal for itself [1]. In most realistic cases, therefore, the raw data, or the means of obtaining them, will be widely available before their information content is understood. The blc1 candidate, news of which reached the press while its analysis was still in progress and which was ultimately shown to be radio-frequency interference [20], illustrates how quickly information about a candidate signal can enter the public domain. The practical question is not how to control access to the data, but how to recognise, communicate, and manage the moment a detection is discovered to be information-rich.

An information-rich detection would also widen the range of actors and disciplines involved. The verification of a candidate signal is principally a task for astronomers and instrument specialists. The decoding of an information-rich signal is not; it would engage cryptographers, information theorists, linguists, and computer scientists, and would rely heavily on machine-learning methods. No doubt astronomers, a community rarely reluctant to extend its skills to problems well outside its own knowledge domain, would also have a go themselves, as would skilled amateurs. Here a distinction must be drawn that is easily missed. Machine-learning methods are already effective at the recognition of modulation and structure in unfamiliar radio signals [21,22], and this is a tractable, well-posed problem. The semantic interpretation of recovered content is not the same problem, and success at the former does not imply competence at the latter. Systems trained exclusively on human languages, mathematics, and imagery may impose familiar interpretations on data that do not warrant them, producing outputs that are fluent and internally coherent yet ultimately unfaithful to the signal [23]. The apparent success of such a system may reflect the strength of its priors rather than the structure of the received data. The concern is not confined to a failure to decode; it extends to a confident but erroneous decoding mistaken for a genuine one, a difficulty compounded by the likelihood, already evident in controlled exercises, that a genuine message may admit more than one internally coherent interpretation [7]. Where recovered content makes empirical claims, competing interpretations could in principle be tested against nature, but testable content might take years to verify.

Risk, too, takes a different form. A simple carrier is inert. A body of recoverable content is not necessarily so. This does not depend on any assumption of hostile intent on the part of a transmitting civilisation; the more immediate considerations are terrestrial, and include the misinterpretation of benign information, action taken on an erroneous decoding, and in the more speculative case of machine-readable content, the execution of instructions that are not understood. McConnell anticipated that a message might carry software rather than static content, in effect what would now be called a large language model developed by another civilisation, as valuable as it would be unpredictable once run [24]. The possibility that a complex message might carry technical hazards has been argued in its strongest form by Hippke and Learned, who contend that a message requiring computation to analyse it cannot be decontaminated with certainty and might, in principle, pose a serious risk [25]. The position taken here is more measured and concerns prudent stewardship rather than the counsel of non-reception that such arguments can invite; the general typology of information hazards [26] offers a more proportionate frame than any presumption that a message should be left unread. It must also be emphasised that entropy is an ambiguous indicator. Very high entropy is characteristic of random or unstructured processes, but also of highly compressed or encrypted data, which may be practically indistinguishable from noise [27]. An uncompressed, information-bearing message would instead be expected to display detectable structure and an intermediate entropy signature [5]. Apparent structure alone is nevertheless not evidence of a message, since propagation and instrumental effects may also generate it. The relevant point is that information, unlike a pure carrier signal, is the kind of entity that can be applied, and therefore the kind of entity whose handling repays forethought. Taken together, these considerations indicate that an information-rich detection would not simply represent a more elaborate instance of a beacon detection, but a situation of a different character, with distinct scientific, institutional, and societal consequences.

## 4. Competitive decoding and the incentive to exploit

Among the features of an information-rich detection that supplementary guidance would need to weigh is the pattern of behaviour such a detection is likely to provoke. If it came to be widely believed that a verified extraterrestrial signal contained knowledge capable of being extracted and applied profitably, a competitive effort to acquire, decode, and exploit that signal would be a plausible and arguably predictable outcome. This expectation follows not from any particular pessimism about human nature but from the ordinary behaviour of actors in relation to a resource that is both valuable and scarce.

States constitute the most obvious such actors, and the risk they pose is already recognised. Wisian and Traphagan argue that even a passive detection may prompt state-level attempts to secure a communication monopoly, with a corresponding risk of international conflict [15]. It is worth being precise about the object of competition, because this is where an information-rich detection departs from the case the existing geopolitical literature has principally examined. Wright, Haramia, and Swiney offer good reasons to doubt that a state could monopolise a *communication channel* with an extraterrestrial intelligence, and on that basis favour transparency over securitisation [16]. A recoverable informational *payload* (to borrow the engineering term for the content a transmission carries) is a different object. It need not be monopolised at the point of collection to be valuable, and it may be received independently by many parties; the competition it invites is not to control a conversation but to be first to decode and to apply the knowledge received. The reassurance that a channel cannot be monopolised does not, therefore, remove the incentive to exploit a payload, and it is this latter incentive that an information-rich detection would bring to the fore.

It would be a notable omission, moreover, to consider states alone. Private commercial actors are in several respects well positioned to pursue an information-rich signal and would have substantial incentives to do so. The underlying resource is not readily monopolised at the point of collection: a sufficiently strong signal can be detected by any adequately equipped facility. Some large corporations possess many of the assets required for such an effort, including high-performance computing, advanced AI capabilities and signal-processing expertise, cloud-scale data infrastructure, radio-frequency engineering, and, in some cases, extensive ground-station networks. This makes them very plausible participants in any competitive attempt to detect, analyse, and decode the signal. An organisation able to recover even a fragment of an extraterrestrial data stream bearing on energy, materials, or computation might regard the result as possessing extraordinary commercial value. The *A Sign in Space* experience noted in Section 2 suggests that such interest need not wait for a real detection [8]. The existing literature, in concentrating on state actors, has largely left this dimension unexamined. It is a natural candidate for supplementary attention, for an information-rich detection would, for the first time, introduce the prospect of a substantial commercial prize within SETI, and the presence of such a prize would tend to elicit private efforts to secure it, with no obligation of transparency or disclosure.

These incentives would generate pressures worth anticipating. A competition to be first to claim a decoding would reward speed at the expense of rigour and would increase the likelihood of premature, erroneous, or self-interested interpretations entering circulation. An incentive to withhold methods and results would run counter to the collaborative verification on which scientific reliability depends. And the surrounding public environment would present a further challenge: in the contemporary information ecosystem, a confirmed detection would be followed rapidly by fabricated decoding, speculative claims, and deliberate misinformation, propagating more quickly than careful analysis could respond [28,29]. These are challenges the declaration's strengthened communication guidance already recognises, and which an information-rich detection would surely intensify. The determination of what a signal actually contained would become entangled with a contest over public credibility, for which the *A Sign in Space* experience offers a useful precedent [7].

It should also be noted that exploitation would not, of course, be solely confined to the information-rich case. Even a monotone beacon would create opportunities of its own, independent of anything the signal contains. Commercial interests could include the attention economy, in which media and online platforms profit from engagement regardless of accuracy; the sale of privately funded follow-up observations, data products, or privileged access; speculative activity in financial and digital-asset markets reacting to the news; and commercial or self-appointed attempts to transmit a reply on behalf of humanity.

All of this suggests that an information-rich detection would arise within an environment of competing states, very capable commercial actors, powerful but not fully understood AI systems, and rapid misinformation. This is without doubt an environment where supplementary guidance is very much warranted.

## 5. Upholding openness under new pressures

The considerations above bear on the principle most central to SETI post-detection practice: the commitment to openness and transparency, which the 2026 declaration enshrines in its encouragement to preserve and disseminate the data, methods, and code underlying a detection [1]. This commitment is not incidental, and there is no case for weakening it. Open and independent access to detection data is the mechanism by which instrumental error, local interference, and motivated misinterpretation are guarded against, and by which public confidence in a claimed detection is established. It is difficult to conceive of a credible SETI enterprise that did not rest upon it, and there is a strong case, advanced by Wright, Haramia, and Swiney among others, that transparency is not only a scientific virtue but the most effective protection against destabilising competition [16]. This paper shares that view without reservation: openness should remain the default, and the risks of secrecy exceed the risks of disclosure.

The question that an information-rich detection would raise is therefore not whether openness should be maintained, but how it can most effectively be upheld and reinforced under conditions the simpler beacon scenario does not present. Two features of a recoverable payload deserve particular attention. The dissemination of such content is irreversible: once copied widely, it cannot be recalled, so decisions about its release carry a weight that the release of a beacon does not. And the value, implications, and any hazards of the content may be imperfectly understood at the moment those decisions must be taken. These are not reasons to depart from openness; they are reasons to consider, in advance, how an open and reproducible scientific process can be conducted responsibly when release of the material is irrevocable and its content potentially exploitable. Nor is that question without precedent: discussions of dual-use research have led to the development of governance frameworks that evaluate new scientific activities across multiple dimensions, including proportionality, risk management, evidence, justice, and public accountability. These established approaches offer practical experience from which supplementary analyses could profitably draw [30].

Openness is a norm that binds only those who choose to observe it. The scientific community's commitment to transparency cannot, by itself, govern the conduct of a state that regards a signal as a matter of national security, or of a corporation that regards it as a commercial asset; neither is a party to the protocols. The boundary is, in any case, not a sharp one: much of the contemporary search is funded philanthropically or carried out in partnership with private organisations on publicly owned facilities, and some of these programmes, for example, Breakthrough Listen's open-data practice, have made openness a fundamental part of how they operate. The partnerships involved are already diverse. Breakthrough Listen, funded philanthropically through the Breakthrough Initiatives, conducts its searches on publicly operated facilities including the Green Bank and Parkes telescopes [31]; the Allen Telescope Array was built with substantial private philanthropic and corporate support [32]; and citizen science has long been part of the enterprise, from the millions of volunteers who contributed computing time to SETI@home [33] to the global community that extracted and decoded the *A Sign in Space* message [7,8]. These partnerships have been highly productive and, in the main, committed to openness. They nevertheless show that the actors likely to be involved in a real detection, whether of a beacon or of an information-rich signal, already span public, philanthropic, commercial, and citizen sectors, each with different obligations, incentives, and expectations of credit. Supplementary guidance would need to consider how the declaration's commitments apply across this mixed funding and ownership landscape. In an environment in which powerful actors possess independent means of reception and analysis, the community's openness is best understood not as a guarantee that all parties will be open, but as the foundation of its own credibility and authority. That authority, the trust that attaches to a transparent, reproducible, well-communicated scientific account, may in fact be the community's most effective instrument in a contested environment, and future contributions could valuably consider how it is preserved and strengthened. The realistic aim is not to compel universal openness, which no protocol can achieve, but to ensure that the open scientific record remains the authoritative reference against which unilateral or opportunistic claims are measured [34]. How best to secure that outcome, when content is exploitable and not every actor is a willing participant, is a question well suited to the supplementary development of the current post-detection framework.

## 6. Toward supplementary contributions

The preceding sections are directed at a single, constructive proposition: that the detection of an information-rich technosignature is a scenario sufficiently distinct in its demands to merit dedicated consideration as the community develops supplementary guidance in support of the 2026 declaration. Consistent with that aim, this paper does not advance revisions to the adopted protocols, nor does it advocate specific measures. It instead seeks to identify the scenario and offer a preliminary set of questions that supplementary work can address.

Several questions follow. A beacon effectively conflates two decisions: confirming that a signal is extraterrestrial and deciding how its content should be handled. An information-rich detection may require these questions to be considered separately. If so, how can this be done while maintaining the declaration's commitment to openness? Given that information content may be recognised progressively rather than at a single moment, how should that transition be identified and communicated, and what interim practices are appropriate while it is under way? How can the declaration's commitment to openness best be upheld and reinforced when recoverable content is potentially exploitable and when powerful actors outside the SETI community may not share that commitment? What standards of evidence and reproducibility ought to govern claimed decoding results, particularly those produced with the assistance of AI whose outputs may be persuasive but neither fully traceable nor demonstrably faithful to the underlying data? How might the community best respond to competing and possibly incompatible interpretations of the same signal, and to the misinformation that would accompany them? And by what means might the wider public interest be served in the face of possible monopolisation or commercial appropriation of extraterrestrial knowledge? Scenario-based and strategic-foresight methods [11,12,13] offer a practical means of exploring these questions, and the behaviour of the different actors involved, before a detection occurs.

These questions are offered not as an agenda to be resolved here but as a starting point for the supplementary development that the declaration anticipates and that the standing post-detection function and the wider readiness effort are well placed to carry forward [2,9,10,11]. The protocols are designed to evolve as circumstances require [1,2]; the argument of this paper is simply that the information-rich scenario, and its implications for the responsible practice of openness, warrant explicit attention within that ongoing work. The advantage of considering the scenario now lies in the opportunity for unhurried reflection, at a time when no actual detection is compressing the interval available for it.

## 7. Concluding remarks

A signal from an extraterrestrial civilisation might take the form of a beacon whose entire content is the demonstration of its own existence, or of a transmission bearing recoverable and potentially exploitable information. The 2026 declaration provides the foundation for the verification, preservation, announcement, and communication of a detection in either case, and it explicitly addresses information content, requiring that it be recorded and made publicly available along with the underlying data and methods. As a living instrument that anticipates supplementary best-practice guidance, it also provides the mechanism by which particular scenarios can be examined in greater depth as the field develops.

This paper has sought to identify one such scenario. An information-rich detection would shift the post-detection effort from the verification of a signal towards the stewardship of the information it contains; it would extend the relevant timescale, widen the range of actors and disciplines involved, and introduce competitive incentives on the part of states and of commercial organisations alike to decode and to exploit the signal. Because the information content of a signal may be recognised only gradually, much of the relevant data could already be openly available by the time that content is understood. These features do not call the adopted protocols into question, but they do indicate the value of supplementary guidance addressing how the community's commitment to openness is best upheld and reinforced under new pressures. The scenario is offered here in that constructive spirit, as a subject for the continuing development of the framework the community has established. The time to consider what an information-rich detection would ask of us is before such a detection occurs, and not after.

### Declaration of competing interest

The author chairs the SETI Committee of the International Academy of Astronautics and led the task group that prepared the 2026 update to the Declaration of Principles discussed in this paper.

### Acknowledgements

I thank the referees for their careful reading of the manuscript and for their constructive comments, which have improved the paper.